# Templating Shuffles

Qizhen Zhang[1,2,3], Jiacheng Wu[4], Ang Chen[5], Vincent Liu[1], Boon Thau Loo[1]
[1]*University of Pennsylvania,* [2]*Microsoft Research,* [3]*University of Toronto,* [4]*University of Washington,* [5]*Rice University*
[1]{qizhen, liuv, boonloo}@seas.upenn.edu, [2]qizhenzhang@microsoft.com, [3]qz@cs.toronto.edu
[4]jcwu22@cs.washington.edu, [5]angchen@rice.edu

## ABSTRACT

Cloud data centers are evolving fast. At the same time, today's large-scale data analytics applications require non-trivial performance tuning that is often specific to the applications, workloads, and data center infrastructure. We propose TeShu, which makes network shuffling an extensible unified service layer common to all data analytics. Since an optimal shuffle depends on a myriad of factors, TeShu introduces *parameterized shuffle templates*, instantiated by accurate and efficient sampling that enables TeShu to dynamically adapt to different application workloads and data center layouts. Our preliminary experimental results show that TeShu efficiently enables shuffling optimizations that improve performance and adapt to a variety of data center network scenarios.

## 1 INTRODUCTION

Large-scale data analytics systems [7, 10, 13, 15, 16, 24, 36, 47] are a key application class in modern data centers. Universal to these platforms is the need to transfer data between blocks of compute. Broadly defined, processing typically occurs in a few key phases (Figure 1): (i) **compute**, in which workers independently process their local shard of data, (ii) **combine**, optional, in which preliminary results are locally processed to reduce the data that passes through (iii) **shuffle**, the process of resharding and transmitting data to the next phase of compute.

Of particular note in this pipeline is the shuffle phase. Compression, serialization, message processing, and transmission all contribute to the CPU, bandwidth, and latency overhead of this phase. More so than the other two phases, shuffles, if planned poorly, can become a significant throughput and latency bottleneck of the system. Application performance is often gated on tail completion time of the shuffle. Because of that, there is a rich history of work in tuning the behavior of this phase [15, 21, 28, 29, 48]. More recently, shuffle has been shown to be a major performance bottleneck in data analytics on emerging cloud platforms [27, 32, 34].

Unfortunately, this tuning process is non-trivial as performance characteristics are not only highly dependent on the workloads, but also on the underlying data center architecture. To make matters worse, more and more big data systems opt for disaggregated in-memory and virtual disk storage that cross a network where interactions are complex [4, 14, 33], the topology is constantly changing due to failures [6, 9, 17, 22], and next-generation designs are increasingly sophisticated [3, 37, 44, 45]. Prior work has resulted in complex solutions that either fail to provide portable performance [20, 23, 29–31] or are difficult to reason about [5].

This paper proposes a novel solution: a templated shuffle (TeShu) layer that can adapt to application data and data center infrastructure, on top of which existing and future systems can be implemented.

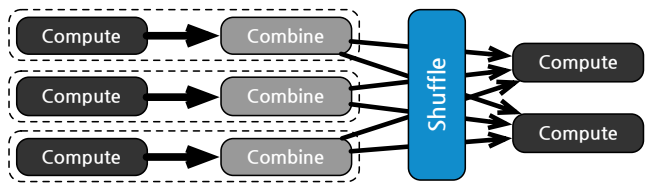

**Figure 1: The general structure of most data analytics systems: computation, combination, and shuffling. This process can be applied to various systems like MapReduce, graph processing, and cloud-based analytics systems.**

The design of TeShu centers around *parameterized shuffle templates*, which provide a set of shuffle primitives that can greatly simplify the job of writing performant data analytics software. TeShu engineers can write a wide array of shuffle templates. As they may not know the characteristics of the infrastructure or workload a priori, they instead leave various parameters undefined. At runtime, TeShu instantiates the shuffle template by populating the parameters using knowledge of the underlying topology and data achieved via sampling.

TeShu enables infrastructure-aware optimizations that provide the illusion of hand-tuned performance, but in a portable fashion where a programmer could deploy graph systems, e.g., Pregel or Spark jobs, without worrying about how shuffling (and hence overall performance) is impacted by the workload characteristics, network topologies, and failure scenarios. Our contributions follow.

- **Customizable shuffle with templates**. We present the design of TeShu and demonstrate how its *shuffle templates* are expressive enough to support a wide range of big data analytics systems and shuffle optimizations.
- **Network-aware shuffling.** We demonstrate how infrastructure-level optimizations are made possible by instantiating shuffle templates at runtime. In particular, we present an adaptive optimization that dynamically chooses the best shuffling strategy for a given data center network topology.
- **Evaluation.** Our evaluations on real-world graph workloads show that TeShu can enable adaptive optimizations that significantly improve application performance. Our proposed sampling-based parameter tuning can achieve high accuracy with low sampling cost.

## 2 TESHU OVERVIEW

TeShu is centered around three core ideas: a common shuffle layer customizable via templates, shuffle optimizations adaptive to workloads and networks, and a sampling mechanism that enables adaptation. Details follow.

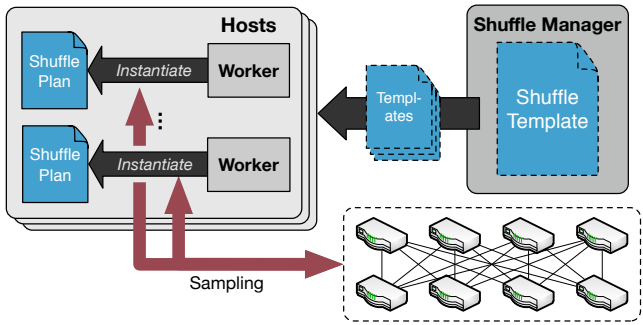

**Figure 2: The architecture of TeShu. When `shuffle` is invoked, a shuffle manager ships shuffle templates to the application.**

*Customizable, templated shuffle as a common layer.* Despite their simplicity, compute, combine, and shuffle can support use cases from graph processing (e.g., Pregel, Giraph) to SQL queries (e.g., SparkSQL and Hive). At its core, shuffle simply transfers data across nodes. The source/destination, transfer rate, and mode of synchrony can vary between systems, but in every case, the interface is consistent. TeShu supports this design by enabling a cleaner layering between applications and infrastructure. Rather than spend time tuning each application, users of TeShu template the shuffle layer that is common to all upper-layer systems.

*Adaptive shuffle.* TeShu can take into account the workload, combiner logic, shuffle pattern of the application, and network topology to adapt the shuffle to the environment. At runtime, TeShu instantiates execution plans and directs the shuffle data for higher layers. We envision that TeShu runs as a service that many big data platforms can invoke.

*Sampling-based adaption.* Finally, we allow shuffles to dynamically adapt to the query workload and network by sampling data that most efficiently tests the efficacy of optimizations. Our application-centric sampling avoids classic constraints that come with the statistical estimation of population parameters such as knowing the distribution. Empirically, we find that testing a small fraction of the data (as low as 0.01% in real-world workloads) already leads to high accuracy.

## 3 CUSTOMIZABLE SHUFFLE

In TeShu (Figure 2), a Shuffle Manager is deployed as a service by the infrastructure provider along with *shuffle templates*. During job execution, the application will invoke the shuffle API, which results in an RPC to the Shuffle Manager, which and application workers cooperate to instantiate the template to form a full shuffle plan. Workers execute the plan to shuffle their data.

### 3.1 Shuffle API

In TeShu, shuffle operations are defined as instances of concurrent communication between a fixed set of sources and destinations. Programs invoke shuffles for a variety of reasons and in a variety of different contexts. These include loading data from network storage to workers, distributing intermediate values between iterations, and aggregating results. All of these uses can be specified using the following abstraction:

```
shuffle(wId, templateId, shuffleId,
    srcs, dsts, bufs, partFunc, combFunc)
```

| Parameter | Type | Description |
| --- | --- | --- |
| wId | $i$ | Worker identifier. |
| templateId | $i$ | Shuffle template identifier. |
| shuffleId | $i$ | Shuffle invocation identifier. |
| srcs | $\{s_1,...,s_n\}$ | The set of identifiers for workers where data resides. |
| dsts | $\{d_1,...,d_m\}$ | The set of identifiers for workers to which data is moved. |
| bufs | $\{D_1,...,D_k\}$ | Buffers for sent or received data. |
| partFunc | $f:(D,\{d\}) \to d_k$ | Data partition function (optional). |
| combFunc | $f:(D_1,D_2) \to D$ | Message combiner function (optional). |

**Table 1: Parameters to the shuffle call.**

In the base case, the RPC of a shuffle invocation requires a worker identifier, a shuffle template identifier specifying which template to use, a shuffle identifier, a list of sources, and a list of destinations of the shuffle operation. For example, in Hadoop, the sources and destinations will be the list of Map and Reduce workers.

Other types of shuffles can be specified using optional parameters. For instance, communication patterns for reduction and aggregation can be implemented with a *partition function*. The function takes each piece of data and maps it to a destination worker. A simple example of a hash-based partition function (the default partition function) is the following:

```
partFunc(D, dsts):
    return hash(D) % dsts.size
```

Finally, the shuffle call can include a commutative and associative *combiner function*. For example, the combiner function for wordcount takes a set of (word, count) tuples and performs aggregation as follows:

```
combFunc((w, n1), (w, n2)):
    return (w, n1+n2)
```

TeShu is agnostic to the type of analytics job. Modifying existing code bases to run shuffling with TeShu is natural.

### 3.2 Shuffle Templates

The result of shuffle calls are specialized shuffle plans that define the communication and processing to be done at each node to execute the larger shuffle operation. System operators do not define shuffle plans directly; instead, they define Python-like shuffle templates with parameters to be filled in, automatically, locally on workers later.

Table 2 lists these parameters (functions). Five of them act as primitives for basic communications (SEND, RECV and FETCH), data partition (PART), and message combine (COMB). Those primitives are easily translated to the language of each system. In addition, SAMP is the sampling function for estimating a particular shuffle cost. These parameters suffice to express a variety of shuffle algorithms including those in Table 3.

We note that these functions as well as the shuffle call are synchronous, meaning that they run to completion in the invocation to ensure that the shuffle logic is executed and the data is delivered. We leave adding asynchronous communication to support overlapping computation and communication as future work.

| Template Parameter | Description |
|---|---|
| SEND(dst, msg) | Send msg to dst. |
| RECV(src) | Return the data received from src. |
| FETCH(src) | Return the data fetched from src. |
| PART(msgs, dsts, partFunc) | Partition msgs into dsts according to partFunc. |
| COMB(msgs, combFunc) | Combine msgs according to combFunc. |
| SAMP(msgs, rate, partFunc) | Sample msgs based on rate and partition function partFunc. |

Table 2: Shuffle template parameters that are automatically instantiated when the template is received by workers.

**Push/pull communication.** To support both *pull* (e.g., MapReduce systems) and *push* (e.g., Pregel-like systems) shuffle patterns, TeShu separates the *sender template* and *receiver template* in a shuffle. SEND and RECV are designed for a push model where senders send messages, and FETCH is designed for a pull model where receivers proactively request messages.

Consider the simple 'vanilla shuffling', as in MapReduce, where sources send messages to a list of destinations. The pull-mode template for this is (*sender template*: call PART(bufs, dsts, partFunc) to partition messages; *receiver template*: for each n in srcs, call bufs[n] = FETCH(n) to fetch messages). More examples are in next section.

**Adaptive optimization.** To support adaptive shuffle optimization, TeShu allows applications to sample messages. The SAMP function takes a set of messages msgs and sampling rate rate, performs *partition-aware sampling* (detailed in the next section) based on partFunc, and returns the sampled messages. Those samples can be used to run small, yet accurate, shuffle experiments to estimate parameters. The use of SAMP includes testing the efficiency of a particular shuffle, and estimating the reduction ratio if a combiner is applied on a set of messages.

### 3.3 Shuffle Management

The shuffle manager serves as a central controller to coordinate template instantiation and execution by application workers. Currently, the primary functionality of the manager is to store and serve templates. System operators first install optimized shuffle templates according to their data center network topology to the shuffle manager. From an application's perspective, the shuffle API looks very similar to today's big data execution model: individual workers will call the shuffle function described above. Senders and receivers can arrive at the shuffle at different times and the data can finish transferring to different destinations.

Specifically, when a worker invokes the shuffle call, and if the requested shuffle template is not cached locally, an RPC operation is issued to the shuffle manager to request the template. Upon receiving an RPC request, the shuffle manager allocates a record in memory for the request with necessary information, e.g., the worker identifier, the shuffle identifier, the template identifier, and current timestamp, to indicate the start of a shuffle at a particular worker. Then it ships the template back to the worker. Once the worker receives the response from the shuffle manager, it continues by (1) populating the template with the arguments of the shuffle invocation, i.e., the parameters in Table 1, (2) compiling the template into a physical shuffle plan, which is an executable for the system, e.g., a native library, (3) caching the template and the plan locally, and finally (4) executing the shuffle plan. Later invocations to the same template directly utilize the cached executable, with an asynchronous RPC request sent to the shuffle manager to record the shuffle.

| Shuffle Algorithm | Description | Pattern | LoC |
|---|---|---|---|
| *Vanilla shuffling* | Send messages from sources to destinations. | Push/Pull | 5 |
| *Coordinated shuffling* [21] | Optimize shuffle bandwidth on NUMA nodes. | Pull | 9 |
| *Bruck shuffling* [38] | Schedule flows to avoid single process bottleneck. | Push | 11 |
| *Two-level exchange* [27] | Group small shuffles to reduce cost in the cloud. | Push | 18 |
| *Network-aware shuffling* | Adaptively shuffle data at data center scale. | Push/Pull | 48 |

Table 3: Examples of shuffling algorithms and optimizations. LoC indicates the number of lines of TeShu template code.

When the shuffle plan is finished by a worker, before shuffle returns, an RPC request indicating the completion of the shuffle is sent to the shuffle manager. The shuffle manager allocates another record to indicate the end of the shuffle. It can leverage these records to track the progress of each worker for a shuffle operation to handle stragglers or log the records to facilitate fault tolerance. The shuffle manager can also be replicated and sharded for fault tolerance and scalability. We leave these investigations to future work.

## 4 EXPRESSIVENESS

The parameterized shuffle templates in TeShu can support a wide range of shuffle algorithms. We now describe several optimized algorithms that are proposed recently, and briefly show how they can be expressed in TeShu. We focus on an optimization for data center infrastructure, which we term *network-aware shuffling*, that can significantly improve shuffle performance. We further describe how SAMP ensures that our optimization never does worse than the baseline.

Table 3 lists three existing shuffle optimizations: *coordinated shuffling* [21], *Bruck all-to-all shuffling* [38], and *two-level exchange* [27] that we implemented using TeShu templates. *Coordinated shuffling* pairs senders and receivers with two rings and rotates the rings clockwise to maximize the bandwidth for a NUMA machine; *Bruck all-to-all shuffling* schedules flows in an all-to-all pattern to make sure that the shuffle is never blocked in a single process; *Two-level exchange* optimizes data shuffling between serverless cloud functions. It reduces the complexity of all-to-all shuffle on file requests to the cloud storage (quadratic in the number of workers) by grouping workers so that the requests from the workers in the same group can be merged. These three optimizations can be implemented with 9, 11, 18 lines of TeShu template code respectively. We now detail *network-aware shuffling*, an adaptive optimization enabled by TeShu.

```
1   COMB(bufs, combFunc)
2   sNbrs = $FIND_NBRS_PER_SERVER(wId, srcs)
3   sSampMsgs = SAMP(bufs, $RATE, partFunc)
4   (S_EFF, S_COST) = $COMPUTE_EFF_COST(sSampMsgs)
5   if S_EFF > S_COST:
6       sPartMsgs = PART(bufs, sNbrs, partFunc)
7       for n in sNbrs:
8           SEND(n, sPartMsgs[n])
9           sPartMsgs[n] = RECV(n)
10      bufs = COMB(sPartMsgs, combFunc)
11  rNbrs = $FIND_NBRS_PER_RACK(wId, srcs)
12  rSampMsgs = SAMP(bufs, $RATE, partFunc)
13  (R_EFF, R_COST) = $COMPUTE_EFF_COST(rSampMsgs)
14  if R_EFF > R_COST:
15      rPartMsgs = PART(bufs, rNbrs, partFunc)
16      for n in rNbrs:
17          SEND(n, rPartMsgs[n])
18          rPartMsgs[n] = RECV(n)
19      bufs = COMB(rPartMsgs, combFunc)
20  partMsgs = PART(bufs, dsts, partFunc)
21  for d in dsts:
22      SEND(d, partMsgs[d])
```

Figure 3: Network-aware shuffling template in TeShu.

### 4.1 Adaptive Shuffling

Network-aware shuffling optimizes for multi-layer data center networks, starting from worker-level, then server-level, rack-level, and finally global shuffling. At each layer, it combines messages to reduce communication in the over-subscribed data center network, and it leverages sampling to control the potential overhead. We describe its implementation below.

Figure 3 shows the sender template for this strategy in a leaf-spine data center topology, where servers are connected by ToR or 'leaf' switches, and leaves are then connected by a second layer of switches called the 'spine'. Our approach also applies to larger networks.

There are three potential stages to hierarchical shuffling in these types of networks. Before the shuffling begins, each worker performs a local combine operation to reduce the number of messages involved in the shuffling (line 1). The first stage is a server-local shuffle in which workers running on the same physical machine perform local shuffle and combine. Specifically, lines 2 finds the source workers that reside on the same server ($FIND_NBRS_PER_SERVER abbreviates the actual code). Line 3 calls SAMP to sample a $RATE of bufs, which is sSampMsgs, and then we run a shuffle between sNbrs on the sampled messages and applies combFunc to merge the shuffled messages to estimate the data reduction, based on which we estimate S_EFF, the time saved by the reduced data, and S_COST, the time of performing the server-level shuffle ($COMPUTE_EFF_COST abbreviates the actual code). Line 6 partitions the messages for the shuffle between sNbrs. partFunc denotes the partitioning function. Lines 7-9 shuffle the messages, and line 10 applies combFunc to merge messages and replaces bufs with new messages. By this step, all messages that have the same keys in the same server are guaranteed to be combined. This strategy significantly decreases network traffic when there is at least one co-located worker and the combiner is effective on the dataset.

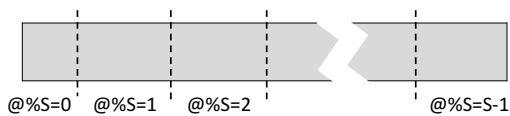

Figure 4: Partition-aware sampling.

The second step is done at a rack level (lines 11-19). Particularly for data centers with high degrees of over-subscription, inter-rack communication can be more costly than communication within a rack. In those situations, reducing the number of messages that are sent across racks can significantly speed up the communication and improve system performance. Finally, the normal global shuffle is performed with the remaining pre-combined data (lines 20-22). The receiver template simply receives data from sources as (for each n in srcs, call bufs[n] = RECV(n) to receive messages)).

**Partition-aware Sampling.** Offsetting the performance benefits of hierarchical shuffling is the overhead of the local combination steps. Network-aware shuffling adaptively applies the local combines based on runtime decisions. It compares the efficiency and the cost (e.g., S_EFF and S_COST at server level). The actual shuffle is executed only if the efficiency is greater. We now describe how the sampling in SAMP works.

A naïve approach is to sample uniformly at random. Unfortunately, we find that random sampling does not work well in practice (see Section 5). Instead, TeShu uses *partition-aware sampling*, which uses consistent hashing to sample the dataset more efficiently. To illustrate this technique, imagine a 'letter count' application that counts the frequency of letters (a-z) in a document. Rather than test a tuple-combiner on a random selection of tuples from random nodes (e.g., $(h,1)$, $(v,1)$, $(z,1)$, ...), a much more efficient method would be to sample the frequency of tuples by the letter (destination). More formally, we use a number $S$, derived from sampling rate, to divide the message destination space into groups from $0$ to $S-1$. Each message on each worker is classified into the $S$ groups using the shuffle's partitioning method so that messages for the same destination are in the same group, as shown in Figure 4 (@ denotes message destination). Finally, we sample messages from a random group $j$ by sending that group from each worker to a worker that acts as the sampling server for evaluation.

## 5 EVALUATION

Our testbed has two racks of 10 servers with both the inter- and intra-rack bandwidths of 10 Gbps. Each server has an Intel E5-2660 CPU with 16 cores at 2.6 GHz, 128 GB RAM, a 10 Gbps network interface card, and 64-bit Ubuntu 16.04 OS. As a preliminary evaluation, we adopted an open-source version of Pregel [2] to test the feasibility and efficiency of shuffle optimizations enabled by TeShu. We use PageRank (PR), and single source shortest path (SSSP), over two real-world graphs with billions of edges: a web graph UK-Web (UK, 3.7B), and a social graph Friendster (FR, 3.6B). We evaluate TeShu's sampling effectiveness, the benefits of network-aware shuffling, and its generality over network and workload variations.

### 5.1 Sampling Performance

The performance of TeShu depends critically on SAMP because high sampling rates can incur significant overhead to shuffle plan execution. Therefore, we first evaluate the accuracy and efficiency of TeShu's sampling algorithm with duplication estimation, which then determines the data reduction rate.

| Sampl. Rate | Ground truth | Rand. | Part.-aware |
|---|---|---|---|
| 0.9 | 0.1833 | 0.1986 | 0.1833 |
| 0.1 | 0.1833 | 0.7241 | 0.1833 |
| 0.01 | 0.1833 | 0.9622 | 0.1832 |
| 0.001 | 0.1833 | 0.9965 | 0.1829 |
| 0.0001 | 0.1833 | 0.9997 | 0.1838 |

Figure 5: Partition-aware sampling significantly outperforms random sampling.

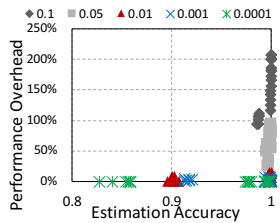

Figure 6: Accuracy vs. overhead of part.-aware sampling.

|  | PR-UK | PR-FR | SSSP-UK | SSSP-FR |
|---|---|---|---|---|
| *Oversubscription Ratio = 10:1* | | | | |
| Execution Speedup | 14.7× | 9.4× | 6.1× | 7.1× |
| Communication Saving | 87.7% | 89.8% | 84.6% | 84.3% |
| Shuffle Decision | S,R,G | S,R,G | S,R,G | S,R,G |
| *Oversubscription Ratio = 4:1* | | | | |
| Execution Speedup | 9.4× | 5.6× | 5.2× | 6.2× |
| Communication Saving | 85.5% | 85.9% | 81.6% | 79.3% |
| Shuffle Decision | S,R,G | S,R,G | S,R,G | S,R,G |
| *Oversubscription Ratio = 1:1* | | | | |
| Execution Speedup | 7.7× | 3.9× | 4.8× | 4.8× |
| Communication Saving | 80.7% | 76.4% | 75% | 66.8% |
| Shuffle Decision | S,G | S,G | S,G | S,G |

Table 4: Evaluation with oversubscription ratios. *S*: server-level shuffle, *R*: rack-level shuffle, *G*: global shuffle.

Table 5 compares the effectiveness of a) random sampling and b) partition-aware sampling for data reduction ratio estimation over a typical workload. Random sampling is close to the true ratio only when the sampling rate is as high as 90%. The overhead of this level of sampling overwhelms any potential improvements. By contrast, partition-aware sampling can achieve very high accuracy even when workers only send 0.01% of their messages for the sample run.

Figure 6 shows the tradeoff between sampling accuracy and overhead, where we vary the sampling rate (from 0.1 to 0.0001) across all applications and datasets. When sampling rate is larger than 5%, the overhead is large—a sampling rate of 10% causes 3× performance slowdown, prohibitively high in practice. As the sampling rate decreases, the overhead drops significantly. Partition-aware sampling achieves excellent accuracy. With a rate of 0.01%, the accuracy is still as high as 80%. Therefore, the rate we have used in the evaluations is between 1% – 0.01%, which achieves 90%+ accuracy with only 8%- overhead. The takeaway is that the SAMP procedure in TeShu is effective and efficient across scenarios.

## 5.2 Adaptive Shuffling Performance

Table 4 shows the execution time speedup achieved by network-aware shuffling across all workloads compared to vanilla shuffling baseline. We note that this optimization can directly benefit any system that is integrated with TeShu, without repeated efforts on optimizing each system.

We observe that when the network is highly oversubscribed (10:1, where inter-rack bandwidth is at a premium), network-aware shuffling can save 80%+ communication cost, and achieve execution speedup from 6.1× to 14.7×. In less oversubscribed environments, it still reduces communication by 66.8–85.9%, and improves performance by 3.9–9.4×.

Table 4 also shows the shuffle strategies decided by network-aware shuffling according to its sampled runs. We observe that when the network is oversubscribed (e.g., 10:1 and 4:1), all three levels of shuffles are performed in the hierarchical shuffling: server level first, then rack level, and finally global shuffling (*S, R, G* in the table). In contrast, when the network is not oversubscribed (1:1), the rack-level shuffling introduces additional overhead. The best strategy is thus server-level, and then global shuffling. The accuracy of SAMP enables this detection, and network-aware shuffling chooses the optimal plan: *S, G* that achieves shorter completion times.

**Robustness to network dynamics.** We additionally evaluated network-aware shuffling with dynamic network scenarios. We injected three random link failures (between ToR and spine switches) for each scenario, and we emulated 100 random failure scenarios. We observe that network-aware shuffling reduces completion times by 5×~8.2×. In fact, with network-aware shuffling, the completion times under failure are very close to those without failures. This shows that network-aware shuffling can dynamically find better strategies and its benefit can be generalized to different network conditions.

## 6 FUTURE DIRECTIONS

**Co-scheduling shuffles.** Currently TeShu schedules individual shuffle invocations for each system. This allows every system to optimize shuffles for their own performance based on the metrics of interest. However, when multiple systems or multiple instances of the same system use TeShu in the same cluster, global scheduling decisions can be important for both performance and fairness. For example, TeShu can identify coflows [14] between shuffle invocations. Scheduling such shuffles together can significantly improve the shuffling performance at application level. Co-scheduling the shuffle calls between multiple systems can also ensure the fair use of the network resources. Enabling flexible shuffle scheduling and identifying the right set of scheduling policies for a specific deployment are the main challenges.

**Handling failures and stragglers.** TeShu currently relies on upper-layer systems to identify failures and stragglers and to restart a shuffle operation. Handling failures of the shuffle manager is relatively easy: we can replicate the management states and shuffle templates installed on the manager. Handling failures of shuffles is challenging as the amounts of data involved in shuffles are often massive. Systems like Spark [47] provide fault tolerance for shuffles by materializing the shuffled data into persistent files. These additional disk activities work fine for shuffles in traditional networks but incur high performance penalty for both large (bottlenecked by bandwidth) and small (bottlenecked by latency) shuffles in emerging fast data center networks. Providing fault-tolerant shuffles with minimal performance overhead for emerging and next-generation cloud networks and making them general for various shuffle templates are worth investigation. Handling stragglers is also challenging. It requires TeShu to have the abilities of tracking the progresses of all shuffle participants and restarting the tasks of a subset of the participants. The shuffle records in the shuffle manager can facilitate these tasks as we discussed in Section 3.3.

**Integrating with in-network techniques.** TeShu currently executes the compute and combine operations in CPUs. Recent years have witnessed many innovations on in-network processing, such as programmable data plane [11, 40] and SmartNICs [26, 41]. TeShu can use these techniques to enable new shuffle optimizations. For example, the `COMB` and `SAMP` functions can be pushed into the network to release the loads from host servers and to gain higher efficiency. How to leverage new network techniques for shuffles and expose them to TeShu's users is another open question.

**Templating shuffles for future cloud data centers.** Data centers evolve fast. Shuffle optimizations that are effective for today's networks may not work for future data centers. For example, hierarchical shuffles [27, 32] for serverless functions that leverage a disaggregated storage backend will be unnecessary if functions can directly communicate [43]. Recent trends on the design of cloud data centers indicate more radical changes. In particular, memory disaggregation [50, 51] separates the computation and main memory for data processing. It translates memory accesses into network communications. Examining the interface of TeShu and developing shuffle templates for this new type of "shuffles" between disaggregated resource pools are a promising direction for future exploration.

## 7 RELATED WORK

*Data analytics optimizations.* Many data analytics systems have been developed [13, 16, 18, 19, 24, 35, 42, 46, 47, 49, 52]. Some also consider optimizing shuffles; but TeShu has a different goal of providing customizability and portability. Recent work has considered ways of understanding the performance of big data systems [29]. Our work instead seeks to abstract away as much as possible to present the simplest possible interface. Prior work has also looked at automatic profiling and adaptation [5]; however, these are typically done across repeated invocations, rather than within a communication round.

*Shuffle optimizations.* A line of recent research investigates the opportunities of optimizing the shuffle layer for improving overall system performance. Camdoop [15] proposes to use direct-connect topologies and in-network aggregation for reducing network traffic for data-intensive applications, while others optimize shuffling for different scenarios, including NUMA [12, 21] and serverless computing [27, 32, 34].

*Network-aware Optimization.* GraphRex [49] proposes data center network-centric optimizations for relational operators in large-scale graph processing. DFI [39] presents a data-flow library that is optimized for RDMA networks. [8] also discusses the importance of modeling the network in parallel data processing. TeShu's position is a general shuffle layer for all big data systems (emphasizing its general shuffle API, shuffle templates and efficient parameter estimation), and therefore existing network-centric shuffle optimizations can be seen as specific applications of TeShu.

*Shuffle service.* Facebook proposes an optimized shuffle service for Spark [48], and Google also offers shuffle service [1]. Compared to TeShu, those shuffle services lack the generality for various data processing systems and shuffle optimizations.

*Execution template* [25] improves job scheduling in big data systems. In contrast, TeShu focuses on shuffles.

## 8 CONCLUSION

TeShu is a system whose goal is to ensure high shuffle performance without requiring complex performance tuning, and to expose a common shuffle API layer to all big data systems. It does this by populating a pre-defined execution template with efficiently sampled data points in order to find the best shuffle strategy. Our initial results demonstrated that TeShu is promising to enable portability and adaptive shuffle optimizations. Many research challenges associated with TeShu and network shuffles in general remain to be explored.